%% file: LFSMar2022.tex
\documentclass[prl,reprint,showpacs,superscriptaddress,nofootinbib]{revtex4-2}%
\usepackage{graphicx}
\usepackage{dcolumn}
\usepackage{bm}
\usepackage{amsmath,amssymb}
\usepackage{comment}

\begin{document}%

\title{Emergence of a hidden magnetic phase in LaFe$_{11.8}$Si$_{1.2}$ investigated by inelastic neutron scattering as a function of field and temperature}

\author{K. Morrison}
\affiliation{Department of Physics and Centre for the Science of Materials, Loughborough University, Loughborough LE11 3TU}

\author{J. J. Betouras}
\affiliation{Department of Physics and Centre for the Science of Materials, Loughborough University, Loughborough LE11 3TU}

\author{G. Venkat}
\affiliation{Department of Materials Science and Engineering, University of Sheffield, Sheffield, S1 3JD, UK}

\author{R.A. Ewings}
\affiliation{ISIS Pulsed Neutron and Muon Source, STFC Rutherford Appleton Laboratory, Harwell Campus, Didcot, Oxon, OX11 0QX, United Kingdom}

\author{A. Caruana}
\affiliation{ISIS Pulsed Neutron and Muon Source, STFC Rutherford Appleton Laboratory, Harwell Campus, Didcot, Oxon, OX11 0QX, United Kingdom}

\author{K. Skokov}
\affiliation{Technical University Darmstadt, Material Science, Functional Materials, Darmstadt, Germany}

\author{O. Gutfleisch}
\affiliation{Technical University Darmstadt, Material Science, Functional Materials, Darmstadt, Germany}

\author{L.F. Cohen}
\affiliation{The Blackett Laboratory, Imperial College London, South Kensington Campus, London, UK}

\date{\today}

\begin{abstract}
The NaZn$_{13}$ type itinerant magnet LaFe$_{13-x}$Si$_x$ has seen considerable interest due to its unique combination of large magnetocaloric effect and low hysteresis. Here we demonstrate, with a combination of magnetometry, bespoke microcalorimetry and inelastic neutron scattering that this is due to the presence of paramagnetic spin fluctuations, which build up as the critical point is approached. While thermal measurements show significant latent heat independent changes in the heat capacity, inelastic neutron scattering reveals the presence of broad quasielastic scattering that persists above T$_c$, in addition to a finite Q quasielastic peak at Q=0.52 $\AA ^{-1}$ (close to a \{100\} Bragg reflection in this system at Q = 0.54 $\AA ^{-1}$). This finite Q quasielastic peak appears only in the paramagnetic state and when in proximity to the itinerant metamagnetic transition. We associate these observations with a hidden (competing) phase and spin fluctuations close to the transition temperature and magnetic field, that persist across the magnetic transition. 
\end{abstract}
\maketitle
\input{introduction.tex}
\input{experimental.tex}
\input{results.tex}
\input{discussion.tex}

\input{conclusion.tex}

\begin{acknowledgments}
{\it Acknowledgements} This work received funding from the European Community's 7th Framework Programme under grant agreement no. CP-FP 214864-2. and EPSRC grants EP/P002811/1 and  EP/T034351/1. Experiments at the ISIS Neutron and Muon Source were supported by beamtime allocations RB1120323 and RB1320200 from the Science and Technology Facilities Council. The authors acknowledge Dr Keith Yendall and the Loughborough Materials Characterisation Centre for assistance and use of the facilities, and Dr David Voneshen for useful discussions. Supporting data from LET is available at https://doi.org/10.5286/ISIS.E.RB1320200; additional data presented here will be made available via the Loughborough data repository under doi 10.17028/rd.lboro.13693525 .  

\end{acknowledgments}
\bibliography{Library}
\bibstyle{apsrev4-1}
\end{document}

%% file: introduction.tex
{\it Introduction.}
\noindent The LaFe$_{13-x}$Si$_x$ (LFS) system (x$<$1.6) has a first order ferromagnetic (FM), to paramagnetic (PM), transition that is tunable in magnetic field, ending at a (tri)critical point (H$_{crit}$, T$_{crit}$) beyond which it is second order.\cite{Franco2017},\cite{Fujita} In this large family of materials the magnetic phase transition can be easily tailored by changing the Fe content\cite{FujitaLFSFe}, hydrogenation, or by substitution of Fe for other magnetic transition metals (Mn, Co).\cite{Shen2009}\\ 
\indent The application of magnetic materials for magnetic cooling was initiated over 20 years ago by the discovery of large entropy changes in the magneto-structurally coupled Gd$_5$Ge$_2$Si$_2$.\cite{PecharskyADM2001}  Since then exploration has yielded several potential material systems.  The most promising magnetic refrigerants appear to be the itinerant ferromagnets LaFe$_{13-x}$Si$_x$ and Mn$_x$Fe$_{1.95-x}$P$_{1-y}$Si$_y$,\cite{GutfleischPhilTrans2016},\cite{NguyenAdvEn2011} both of which exhibit strong coupling between lattice and spin degrees of freedom.  More recently, magnetocalorics have also been suggested as possible contenders for thermal energy harvesting.\cite{WaskeNatEn2019}\\
\indent One of the key requirements of both these technologies is a material with large adiabatic temperature changes that can be repeatedly cycled in field and temperature. LFS is promising as it exhibits a strong magnetovolume phase transition which is first order,\cite{GutfleischPhilTrans2016} therefore has large associated entropy and adiabatic temperature changes.  Crucially, despite the magnetic transition being strongly first order, once extrinsic contributions  are accounted for,\cite{MooreExtrinsic} there is almost no magnetic or thermal hysteresis (an advantage for cooling applications). This has been demonstrated using microcalorimetry by comparing the entropy change due to latent heat to magnetic hysteresis of various magnetocaloric materials.\cite{MorrisonMetMat2014} Another feature of LFS is that fragmentation can broaden the phase transition with respect to magnetic field and temperature.\cite{GutfleischPhilTrans2016} \\
\indent Richter \textit{et al.}, put forward an explanation for the low hysteresis observed in LFS  in terms of a free energy landscape of several local minima associated with different spin states, separated by low energy barriers.\cite{KuzminPRB2007}  It has also been argued that the reduced hysteresis is a consequence of spin fluctuations lowering (renormalizing) the energy barriers one might normally expect of a first order phase transition.\cite{FujitaJMMM2004},\cite{MorrisonPhilMag} \\
\indent More recently, evidence has started to emerge of PM spin fluctuations in LFS\cite{FujitaAPL2016},\cite{FaskeJPC2020},\cite{Zhang0p5LFS} but a full understanding remains incomplete. In this work we present inelastic neutron scattering (INS) data for LaFe$_{11.8}$Si$_{1.2}$ above and below the Curie temperature, T$_c$, and find that quasi-elastic excitations, appear in the PM state, persisting to higher temperatures. In addition, we observe the emergence of a finite Q peak at Q=0.52 $\AA ^{-1}$ that broadens with increasing magnetic field and temperature, which suggests presence of a hidden phase. We argue that these features are linked to both the existence of PM excitations (spin fluctuations) and wave-vector dependent fluctuations as a result of band-structure effects. \\

%% file: experimental.tex
{\it Experimental setup.} A polycrystalline LaFe$_{11.8}$Si$_{1.2}$ ingot was prepared by arc melting constituent elements and annealing in argon at 1323 K (1596 $^{\circ}$C) for 7 days. Magnetometry was performed on a Quantum Design vibrating sample magnetometer at a field ramp rate of 0.5 T/min.\\
\indent	For the INS measurements \cite{INS}, the sample was contained in an aluminum foil packet sealed in a thin aluminum can with He exchange gas, which was cooled by a closed cycle refrigerator (CCR). For the LET measurements a multi-chopper system was employed to simultaneously obtain data for incident energies of E$_i$ = 0.67 (0.017), 1.02 (0.031), 1.74 (0.069), 3.6 (0.2), and 11.6 (1.15) meV (in brackets is the resolution).\cite{BewleyNucIns2011}\\
\indent X-ray diffraction (XRD) patterns of the films were obtained using a Bruker D2 Phaser with incident wavelength $\Lambda$ = 1.54 $\AA$ and a divergent slit width of 1 mm. The measurements were taken in $\theta$/$\theta$ geometry with the diffracted beam optics composed of 0.5 mm Ni monochromator and 2.5$"$ Soller slit followed by a 1D LYNXEYE$^{TM}$ detector with the detector opening set at 5.85 degrees. Reitveld refinement was done using the FullProf Software.\cite{Fullprof} \\
\indent	AC microcalorimetry was obtained using the technique described in Refs. \cite{Minakov, MiyoshiLH}. Selected samples of the order of micrograms were mounted on a 50x100 $\mu$m heater area of a commercial Xensor (TCG-3880) SiN membrane gauge that was adapted for thermal measurements.This technique is used in adiabatic conditions to determine heat expelled/absorbed due to a first order phase transition ($\Delta$Q$_L$), or under isothermal conditions to determine background changes in the heat capacity ($\Delta$C$_p$) and has been proven capable of separating the first and second order contributions to entropy change.\cite{MorrisonJPD2010, MorrisonLHSO, MorrisonPRB2009, MorrisonPRB2013, MorrisonPRB2011}\\

%% file: results.tex
\begin{figure*}[t!]
\includegraphics[width=1.5\columnwidth]{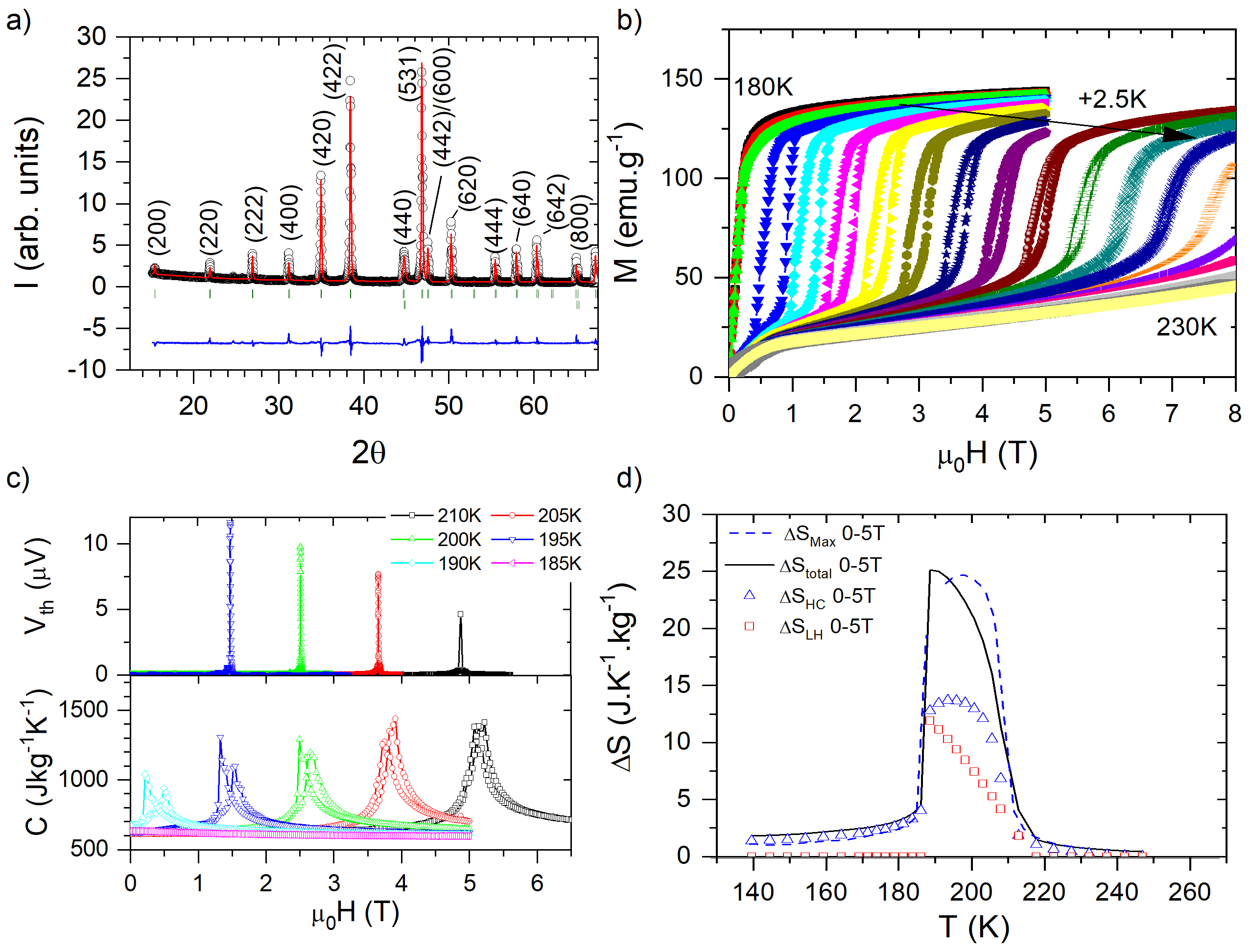}
\caption{Summary of characterisation data for the LaFe$_{11.8}$Si$_{1.2}$ alloy. (a) Powder XRD (symbols) and Reitveld refinement (upper red line) and difference (lower blue line) with peak positions of LaFe$_{11.8}$Si$_{1.2}$ and alpha-Fe phases indicated (green bars). (b) Magnetisation as a function of magnetic field for temperatures ranging from 180 to 230 K at 2.5 K steps. (c) AC microcalorimetry (bottom) and latent heat (top) measurements for various temperatures above T$_c$, measured as a function of magnetic field. (d) Calculated entropy change from magnetisation ($\Delta S_{Max}$) compared to latent heat contribution ($\Delta  S_{LH}$), ac heat capacity contribution ($\Delta S _{HC}$, and total ($\Delta S _{total}$ = $\Delta S_{LH} + \Delta S_{HC}$).}
\label{fig:Fig1}
\end{figure*}

{\it Results.} XRD, shown in Fig. \ref{fig:Fig1}(a) confirmed a majority phase (89.9\%) of NaZn$_{13}$ type LaFe$_{11.8}$Si$_{1.2}$ with lattice parameter 11.474 $\AA$ at room temperature alongside a secondary phase of $\alpha$-iron (10.1\%, a = 2.8648 $\AA$), which is common for this system.\cite{Shen2009} Fig. \ref{fig:Fig1}(b) shows the magnetometry data for this sample which follows the classic itinerant metamagnetic behaviour reported previously.\cite{MorrisonMetMat2014, FujitaJMMM2004} Fig. \ref{fig:Fig1}(c) shows the field driven heat capacity and latent heat measurements, respectively, at various temperatures; the combined entropy change calculated from magnetometry ($\Delta S_{Max}$) and thermal measurements ($\Delta S_{total}$) is shown in Fig. \ref{fig:Fig1}(d). There is documented general agreement between the two methods despite comparing different sized samples.\cite{MorrisonJPD2010, MorrisonPRB2021}\\
\indent Notably the observed latent heat is decreased as the temperature is increased. This was previously attributed to approaching a tri-critical point.\cite{MorrisonPRB2009} As the latent heat decreases, the features observed in the heat capacity measurements initially increase in magnitude, reaching in excess of 200\% of C$_p$(B=0 T, T$>$T$_c$), before decreasing again above the critical point (T* ~ 225 K).\\
\indent Recent work on LFS detected the presence of disordered local moments in the paramagnetic state by resistivity measurements under hydrostatic pressure.\cite{FujitaAPL2016} In contrast, using a fixed spin moment approach, Gruner \textit{et al.}\cite{GrunerPSS} determined that the disappearance of a peak at 27 meV in non-resonant inelastic X-ray spectroscopy (NRIXS) was due to anomalous softening of phonons (thus resulting in large changes to the electronic density of states and the associated lattice entropy). Further observations indicated that the average local moment per Fe atom in part drives the transition from first to second order as it decreases. Finally Faske \textit{et al.}, observed indications of paramagnetic spin fluctuations in LaFe$_{11.4}$Si$_{1.4}$ as diffuse scattering below Q = 0.8 $\AA ^{-1}$ in powder neutron diffraction measurements,\cite{FaskeJPC2020} which was followed by Zhang\textit{ et al.} who observed a quasielastic feature in LFS at Q = 0.5 $\AA ^{-1}$ where x=1.4 that was not present for x=1.8. \\
\begin{figure}[t!]
\includegraphics[width=1\columnwidth]{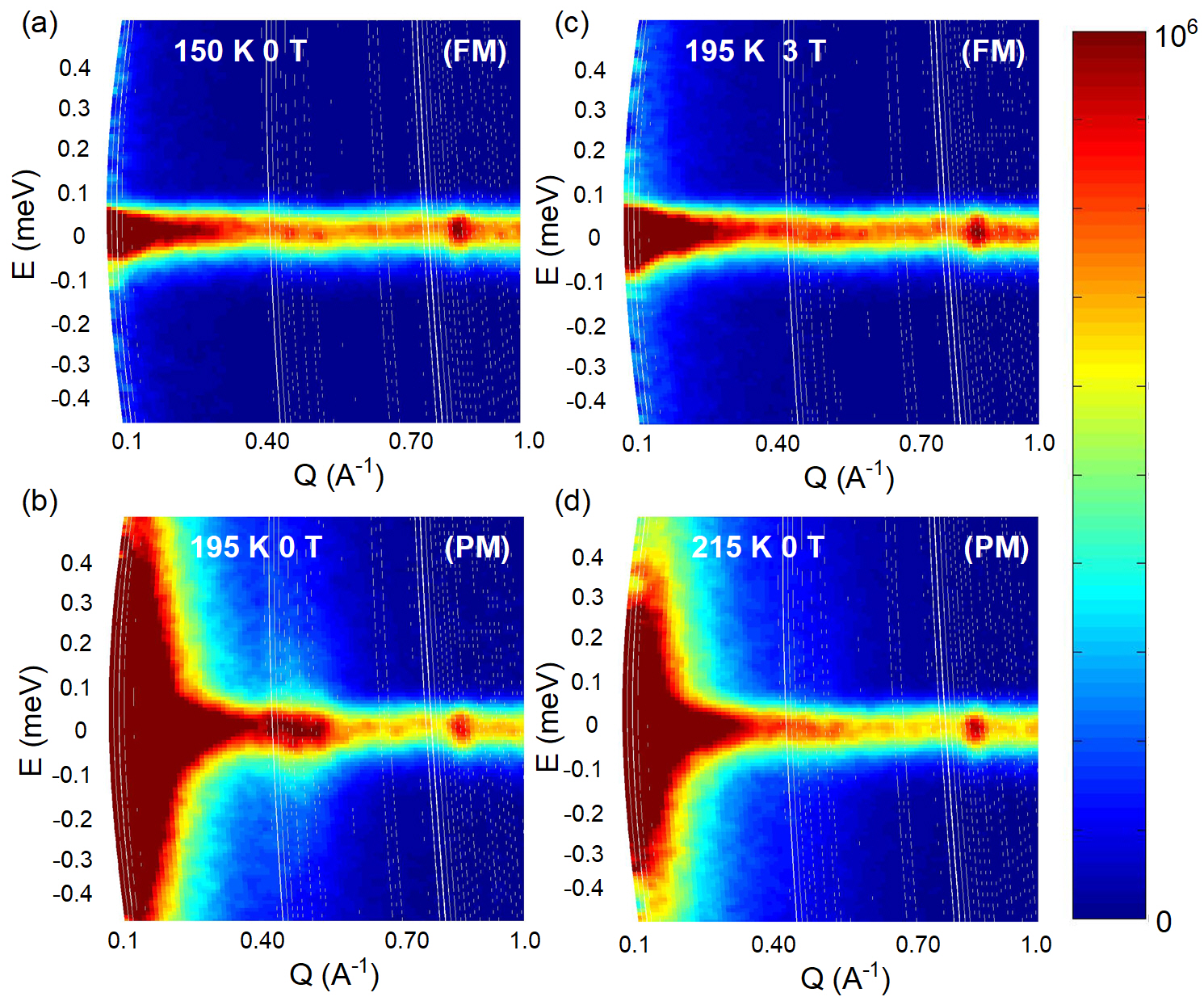}
\caption{Examples of INS of LFS where x=1.2 obtained on LET as a function of temperature and applied magnetic field. (a) Below T$_c$, the sample is in the FM phase. (b) Above T$_c$, giant fluctuations appear about the elastic line from Q = 0 $\AA ^{-1}$ to approximately 0.7 $\AA ^{-1}$ as well as an additional feature at Q ~ 0.52 $\AA ^{-1}$. (c) As the magnetic field is increased and the magnetic phase driven from PM to FM, the fluctuations seen in (b) are suppressed. (d) Close to the tri-critical point (above which the field driven phase transition is continuous), enhanced quasielastic scattering is still present but heavily suppressed and the Q~0.5 $\AA^{-1}$ feature is less evident. }
\label{fig:Fig2}
\end{figure}
\indent INS was used as a tool to explore the possibility of spin fluctuations emerging about T$_c$ that would result in enhancement of the heat capacity and renormalization of the energy barrier. It should be noted that whilst previous measurements  \cite{LanderPRB2018, GrunerPSS, FaskeJPC2020, Zhang0p5LFS} indicated presence of spin fluctuations for Q $\leq$ 0.8 $\AA$ and a phonon peak near 27 meV in the ferromagnetic state, neither had the low Q resolution (to 0.1 $\AA^{-1}$) presented here.   \\ 
\indent Initial measurements of the LaFe$_{11.8}$Si$_{1.2}$ sample performed on MARI showed evidence of a peak in quasielastic scattering close to the zeroth order peak at Q$=$0 $\AA^{-1}$, with a maximum at T$_c$, as shown in the Supplementary Information. This behaviour suggested that the maximum is not due to the expected increase from Debye Waller thermal broadening of the zeroth order peak as the intensity persisted for more than 10 K above T$_c$ ( i.e. well into the paramagnetic state), whilst it dropped off rapidly as the sample approached the tri-critical point (T*).\\
\indent Higher resolution INS was obtained on the LET beamline using a primary incident energy of 3.6 meV. Fig. \ref{fig:Fig2} shows example  E(Q) scans: (a) below T$_c$ (T$_c \approx$192 K), at 150 K, where the sample would have been fully in the FM state; (b) \& (c) just above T$_c$ at 195 K, where in zero field the sample is in the PM state and at 3 T it has been field driven to the FM state; and (d) at 215 K, close to the tri-critical point, T*. Fig. 2(a) \& (c) demonstrate the expected background signal when the sample is in the FM state, where the majority of intensity is confined to the elastic line (E = 0 meV). Fig. 2(b) and (d) show the emergence of quasielastic scattering in the PM state, out to approximately 0.7 A$^{-1}$, in addition to a finite Q peak at $\approx$ 0.52 A$^{-1}$.\\
\begin{figure*}[t!]
\includegraphics[width=2\columnwidth]{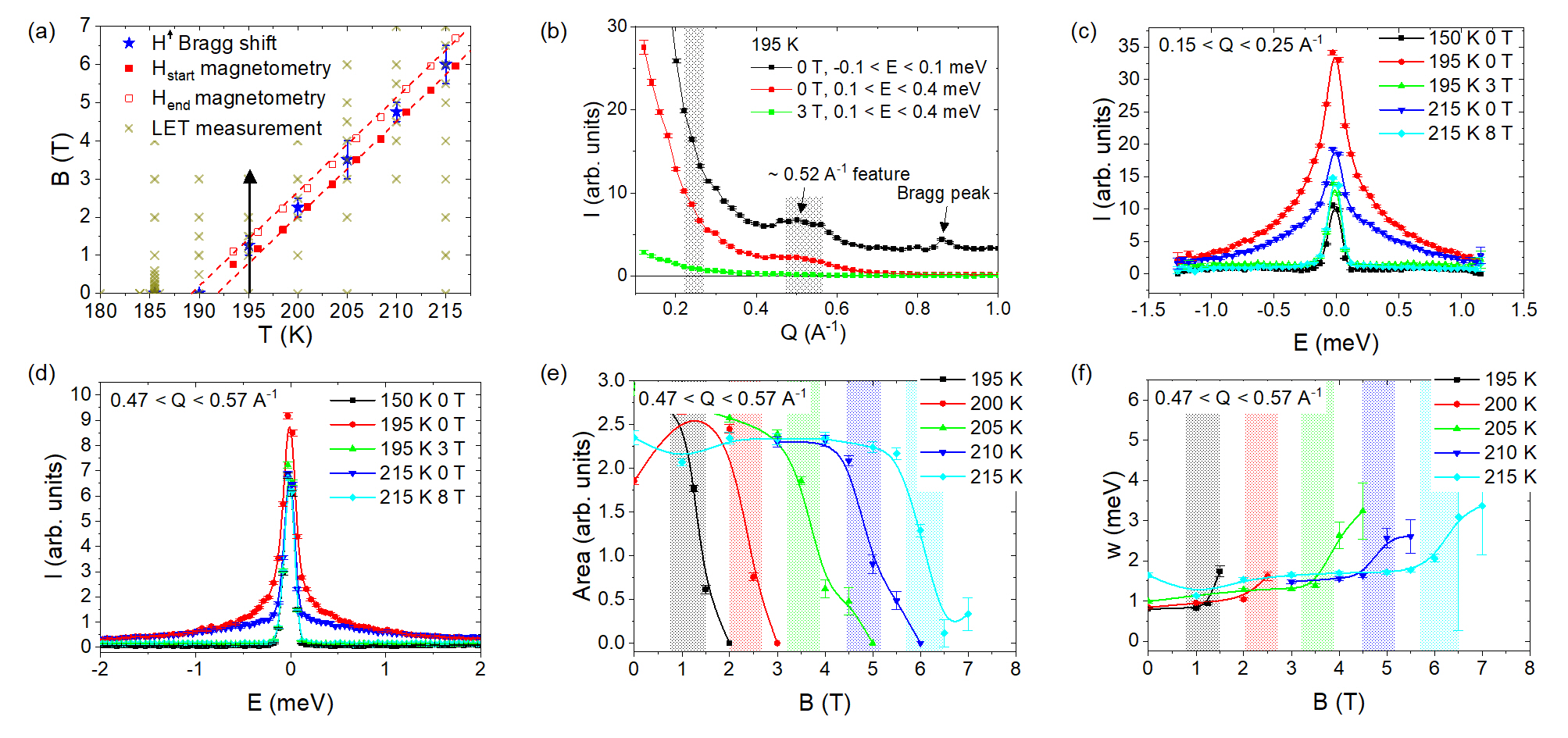}
\caption{Summary of inelastic neutron scattering of x=1.2 obtained on LET, for E$_i$ = {3.6 meV}. (a) Phase diagram of x = 1.2 determined by magnetometry (red symbols) and shift in Bragg peaks at the elastic line (star symbols). The arrow indicates the field history of a measurement, where the magnetic state was reset at B=0 T by heating above T$_c$ and cooling to the chosen measurement temperature. Each available dataset is indicated by the cross symbols. (b) Example linescans just above T$_c$ integrated about -0.1 meV $<$ E $<$ 0.1 meV and 0.1 meV $<$ E $<$ 0.4 meV. The hashed areas indicate the integration window for spin fluctuations and finite Q feature shown in (c) and (d). (c) \& (d) Example linescans above and below T$_c$ for 0.15 $\AA^{-1}$ $<$ Q $<$ 0.25 $\AA^{-1}$ and 0.47 $\AA^{-1}$ $<$ Q $<$ 0.57 $\AA^{-1}$, respectively. (e) \& (f) Results of Lorentzian fit to data shown in panel (d), where corresponding FM background was first subtracted using high field data at the same temperature, and where the field driven transition for each temperature is indicated by the corresponding hatched areas. }
\label{fig:Fig3}
\end{figure*}
\indent Multiple datasets were obtained about the magnetic phase transition driven by field and temperature, as summarized by the phase diagram of Fig. \ref{fig:Fig3}(a). The range of onset fields for a set of fragments is given, which indicated that T$_c$ = 190.5±1.5 K; comparison of heat capacity and neutron scattering data suggests up to a 3 K difference in thermometry of the T$_c$ of the $\mu$g fragment to a mg collection of fragments.  For each set of measurements at a given temperature the magnetic state was first 'reset' by warming the sample in zero field above 195 K. Data was then collected at a given temperature as the magnetic field was increased. Linescans with respect to Q were obtained by integrating I(Q,E) between -0.1$\leq$E$\leq$0.1 meV (elastic line) and 0.1$\leq$E$\leq$0.4 meV (quasielastic region), an example of which is shown in Fig. \ref{fig:Fig3}(b). This demonstrates the presence of diffuse quasielastic scattering away from the elastic line, with decreasing amplitude. Here it can be seen that sharp increase in scattering at 195 K, 0 T, is quickly suppressed by magnetic field as the system is driven to a FM state (3 T). Close to T$_c$, two distinct features were observed:  the first which we attribute to spin fluctuations (SF) for Q $<$ 0.7 $\AA^{-1}$; and the second finite Q peak at approximately 0.47 $<$ Q $<$ 0.57 $\AA^{-1}$.\\
\indent For each measurement where these features were observed, we extracted the corresponding linescan with respect to E (examples in Fig. \ref{fig:Fig3}(c) \& (d)). This demonstrated quasielastic scattering above T$_c$ that decreased in amplitude as the temperature was increased or as the applied magnetic field resulted in a PM-FM phase transition. The diffuse feature at 0.52 $\AA ^{-1}$ (determined by fitting multiple pseudovoigt functions to Fig. \ref{fig:Fig3}(b)) was first normalised, by subtracting the high field data at the same temperature (i.e. expected elastic background in the FM state) and then fit to a Lorentzian function to determine changes in the area, A, and width, w, of this feature with respect to temperature and applied magnetic field. The results of this have been given in Fig. \ref{fig:Fig3}(e) \& (f), where the hatched area indicates the magnetic field across which the PM-FM phase transition occurs at each temperature. It can be seen that the area decreases rapidly on transition from PM to FM phase and the width, w, broadens generally as the temperature and magnetic field is increased. This increase in w makes it more difficult to resolve this finite Q feature at higher temperatures such as in Fig. \ref{fig:Fig2}(d). We note that whilst this feature is close to the position of the \{100\} Bragg peak (Q = 0.54 $\AA^{-1}$), this is a forbidden reflection for this system and is still far enough away that other factors should be properly considered.

%% file: discussion.tex
{\it Discussion.}
In order to explain the observation of strong magnetic fluctuations peaked at finite {\bf Q}, we consider two  phenomenological paths. The first is associated with the magnetic fluctuations that occur with the development of the FM moment. 
Then the total magnetization is written as ${\bf M}_{tot}= {\bf M} + {\bf m}_{||} + {\bf m}_{\perp}$,
where ${\bf M}$ is the ferromagnetic part,  ${\bf m}_{||}$ the longitudinal fluctuations and  ${\bf m}_{\perp}$ the transverse fluctuations.
The minimal Ginzburg-Landau (GL) free energy functional of the system is: 
\begin{eqnarray}
\nonumber
{\cal{F} } = \int d^3x [ &\alpha& {{\bf M}^2_{tot}} + \beta {{\bf M}^4_{tot}}+ \gamma {{\bf M}^6_{tot}}+\delta (\nabla  {\bf M}_{tot})^2  \\
\nonumber
 + &\epsilon& 
(\nabla^2 {\bf M}_{tot})^2 ]
\end{eqnarray}
%
%
%
%
The underlying assumptions regarding fluctuations is that although their spatial average is zero, this is not the case for $\left< m_i^2 \right>$ with $i=\perp, ||$.
As there is no developed long range order, we keep terms up to second order in $m_i$ but it is essential to retain the next higher order in derivatives. The Fourier transform of the part of the fluctuations in the free energy for $T>T_c$ is:
%
%
%
%
%
%
%
%
%
\begin{equation}
\nonumber
{\cal{F}}_{fl}= \int \frac{d^3 q}{(2\pi)^3}  \chi^{-1}(q) \left[m_{||}({\bf q}) m_{||}(-{\bf q}) + m_{\perp}({\bf q}) m_{\perp}(-{\bf q})  \right]
\end{equation}
%
%
\noindent with $\chi^{-1}(q) =  \delta q^2  + \epsilon q^4 + \alpha$.\textbf{}
If $\delta <0$ then $\chi(q)$ is peaked at a finite value of $Q = \sqrt{-\frac{\delta}{2\epsilon}}$. This value is not very sensitive to the value of magnetisation $M$. Both transverse and longitudinal fluctuations need to be retained because the peak is seen in the regime where $M=0$ ($T>T_c$) and drops significantly below $T_c$ where the terms proportional to $M^2$ and $M^4$ are added to $\chi^{-1}(q)$. On the other hand, for the argument to be sound we need to retain a positive susceptibility $\chi(q)$ at all temperatures. This constrains the value of ${\delta}^2/2\epsilon$ to be always less than $\alpha$ for $T>T_c$, and as a result the value of $Q$ should carry a strong temperature-dependence and should decrease to 0 as the temperature approaches $T_c$. As this is not the case, we expect another source of finite-Q fluctuations.

The alternative scenario that appears physically reasonable is that two different phases compete: the pure FM phase and a magnetic phase of finite-Q. The parts of the free energy in the GL functional for each order parameter are independent (with a possible mixing term which does not change the physics qualitatively). In the data, the onset of the winning order parameter at $T_c$ is clearly seen (Fig. \ref{fig:Fig1}), while the second competing phase demonstrates itself only through the observed finite-Q fluctuations (Fig. \ref{fig:Fig2}). These fluctuations are intense in the disordered region, where the phases compete, while they are much weaker below T$_c$, as the prevailing state has already emerged.

 This finite-Q instability (equivalently for $\delta < 0$)  can be attributed  to band structure effects, i.e. possible nesting, as in the case of  Sr$_3$Ru$_2$O$_7$ \cite{Berridge-PRB-2010, Betouras-PRL-2019, FujitaAPL2016}. It is also worth comparing LFS with the effects of quantum fluctuations in ferromagnets. There, quantum fluctuations lead to the onset of a FM  order parameter either through a first order transition or through the development of a finite-Q (e.g. spiral) phase \cite{Belitz-PRL-1999, Chubukov-PRL-2004, Betouras-PRB-2008, Karahasanovic-PRB-2012, Brando-RMP}.
The novel aspect of the present work is that the finite-Q enhancement of the fluctuations is predominantly seen in the PM regime at relatively high temperature (thermal fluctuations) in a 3D material.

%% file: conclusion.tex
{\it Conclusion.}
To summarise, in this letter we present for first time INS measurement of the LaFe$_{11.8}$Si$_{1.2}$ intermetallic as a function of temperature and applied magnetic field. We show emergence of diffuse scattering above T$_c$, which is accompanied by a finite-Q peak at 0.52 $\AA ^{-1}$. The presence of diffuse scattering agrees with previous discussion of the existence of spin fluctuations in LFS, and observations by Gruner \textit{et al.} of diffuse scattering below Q = 0.8 $\AA ^{-1}$.\cite{GrunerPSS} The additional observation of the finite-Q peak in the PM phase, which broadened and decreased in amplitude as the tri-critical point was approached, we argue is the result of competing FM and finite-Q  magnetic phases. As LFS is driven towards a second order phase transition, we would therefore expect the diffuse scattering due to spin fluctuations to remain, whilst the finite-Q peak disappears.

%% file: LFSMar2022.bbl
\begin{thebibliography}{36}%
\makeatletter
\providecommand \@ifxundefined [1]{%
 \@ifx{#1\undefined}
}%
\providecommand \@ifnum [1]{%
 \ifnum #1\expandafter \@firstoftwo
 \else \expandafter \@secondoftwo
 \fi
}%
\providecommand \@ifx [1]{%
 \ifx #1\expandafter \@firstoftwo
 \else \expandafter \@secondoftwo
 \fi
}%
\providecommand \natexlab [1]{#1}%
\providecommand \enquote  [1]{``#1''}%
\providecommand \bibnamefont  [1]{#1}%
\providecommand \bibfnamefont [1]{#1}%
\providecommand \citenamefont [1]{#1}%
\providecommand \href@noop [0]{\@secondoftwo}%
\providecommand \href [0]{\begingroup \@sanitize@url \@href}%
\providecommand \@href[1]{\@@startlink{#1}\@@href}%
\providecommand \@@href[1]{\endgroup#1\@@endlink}%
\providecommand \@sanitize@url [0]{\catcode `\\12\catcode `\$12\catcode
  `\&12\catcode `\#12\catcode `\^12\catcode `\_12\catcode `\%12\relax}%
\providecommand \@@startlink[1]{}%
\providecommand \@@endlink[0]{}%
\providecommand \url  [0]{\begingroup\@sanitize@url \@url }%
\providecommand \@url [1]{\endgroup\@href {#1}{\urlprefix }}%
\providecommand \urlprefix  [0]{URL }%
\providecommand \Eprint [0]{\href }%
\providecommand \doibase [0]{https://doi.org/}%
\providecommand \selectlanguage [0]{\@gobble}%
\providecommand \bibinfo  [0]{\@secondoftwo}%
\providecommand \bibfield  [0]{\@secondoftwo}%
\providecommand \translation [1]{[#1]}%
\providecommand \BibitemOpen [0]{}%
\providecommand \bibitemStop [0]{}%
\providecommand \bibitemNoStop [0]{.\EOS\space}%
\providecommand \EOS [0]{\spacefactor3000\relax}%
\providecommand \BibitemShut  [1]{\csname bibitem#1\endcsname}%
\let\auto@bib@innerbib\@empty
\bibitem [{\citenamefont {Franco}\ \emph {et~al.}(2017)\citenamefont {Franco},
  \citenamefont {Law}, \citenamefont {Conde}, \citenamefont {Brabander},
  \citenamefont {Karpenkov}, \citenamefont {Radulov}, \citenamefont {Skokov},\
  and\ \citenamefont {Gutfleisch}}]{Franco2017}%
  \BibitemOpen
  \bibfield  {author} {\bibinfo {author} {\bibfnamefont {V.}~\bibnamefont
  {Franco}}, \bibinfo {author} {\bibfnamefont {J.~Y.}\ \bibnamefont {Law}},
  \bibinfo {author} {\bibfnamefont {A.}~\bibnamefont {Conde}}, \bibinfo
  {author} {\bibfnamefont {V.}~\bibnamefont {Brabander}}, \bibinfo {author}
  {\bibfnamefont {D.~Y.}\ \bibnamefont {Karpenkov}}, \bibinfo {author}
  {\bibfnamefont {I.}~\bibnamefont {Radulov}}, \bibinfo {author} {\bibfnamefont
  {K.}~\bibnamefont {Skokov}},\ and\ \bibinfo {author} {\bibfnamefont
  {O.}~\bibnamefont {Gutfleisch}},\ }\href@noop {} {\bibfield  {journal}
  {\bibinfo  {journal} {J. Phys. D: Appl. Phys.}\ }\textbf {\bibinfo {volume}
  {50}},\ \bibinfo {pages} {414004} (\bibinfo {year} {2017})}\BibitemShut
  {NoStop}%
\bibitem [{\citenamefont {Fujita}\ \emph {et~al.}(2006)\citenamefont {Fujita},
  \citenamefont {Fukamichi}, \citenamefont {Yamada},\ and\ \citenamefont
  {Goto}}]{Fujita}%
  \BibitemOpen
  \bibfield  {author} {\bibinfo {author} {\bibfnamefont {A.}~\bibnamefont
  {Fujita}}, \bibinfo {author} {\bibfnamefont {K.}~\bibnamefont {Fukamichi}},
  \bibinfo {author} {\bibfnamefont {M.}~\bibnamefont {Yamada}},\ and\ \bibinfo
  {author} {\bibfnamefont {T.}~\bibnamefont {Goto}},\ }\href@noop {} {\bibfield
   {journal} {\bibinfo  {journal} {Phys. Rev. B}\ }\textbf {\bibinfo {volume}
  {73}},\ \bibinfo {pages} {104420} (\bibinfo {year} {2006})}\BibitemShut
  {NoStop}%
\bibitem [{\citenamefont {Fujita}\ and\ \citenamefont
  {Fukamichi}(2005)}]{FujitaLFSFe}%
  \BibitemOpen
  \bibfield  {author} {\bibinfo {author} {\bibfnamefont {A.}~\bibnamefont
  {Fujita}}\ and\ \bibinfo {author} {\bibfnamefont {K.}~\bibnamefont
  {Fukamichi}},\ }\href@noop {} {\bibfield  {journal} {\bibinfo  {journal}
  {IEEE Trans. Magn.}\ }\textbf {\bibinfo {volume} {41}},\ \bibinfo {pages}
  {3490} (\bibinfo {year} {2005})}\BibitemShut {NoStop}%
\bibitem [{\citenamefont {Shen}\ \emph {et~al.}(2009)\citenamefont {Shen},
  \citenamefont {Sun}, \citenamefont {Hu}, \citenamefont {Zhang},\ and\
  \citenamefont {Cheng}}]{Shen2009}%
  \BibitemOpen
  \bibfield  {author} {\bibinfo {author} {\bibfnamefont {B.~G.}\ \bibnamefont
  {Shen}}, \bibinfo {author} {\bibfnamefont {J.~R.}\ \bibnamefont {Sun}},
  \bibinfo {author} {\bibfnamefont {F.~X.}\ \bibnamefont {Hu}}, \bibinfo
  {author} {\bibfnamefont {H.~W.}\ \bibnamefont {Zhang}},\ and\ \bibinfo
  {author} {\bibfnamefont {Z.~H.}\ \bibnamefont {Cheng}},\ }\href@noop {}
  {\bibfield  {journal} {\bibinfo  {journal} {Adv. Mat.}\ }\textbf {\bibinfo
  {volume} {21}},\ \bibinfo {pages} {4545} (\bibinfo {year}
  {2009})}\BibitemShut {NoStop}%
\bibitem [{\citenamefont {Pecharsky}\ and\ \citenamefont
  {Gschneidner~Jr}(2001)}]{PecharskyADM2001}%
  \BibitemOpen
  \bibfield  {author} {\bibinfo {author} {\bibfnamefont {V.~K.}\ \bibnamefont
  {Pecharsky}}\ and\ \bibinfo {author} {\bibfnamefont {K.~A.}\ \bibnamefont
  {Gschneidner~Jr}},\ }\href@noop {} {\bibfield  {journal} {\bibinfo  {journal}
  {Adv. Mat.}\ }\textbf {\bibinfo {volume} {13}},\ \bibinfo {pages} {683}
  (\bibinfo {year} {2001})}\BibitemShut {NoStop}%
\bibitem [{\citenamefont {Gutfleisch}\ \emph {et~al.}(2016)\citenamefont
  {Gutfleisch}, \citenamefont {Gottschall}, \citenamefont {Fries},
  \citenamefont {Benke}, \citenamefont {Radulov}, \citenamefont {Skokov},
  \citenamefont {Wende}, \citenamefont {Gruner}, \citenamefont {Acet},
  \citenamefont {Entel},\ and\ \citenamefont
  {Farle}}]{GutfleischPhilTrans2016}%
  \BibitemOpen
  \bibfield  {author} {\bibinfo {author} {\bibfnamefont {O.}~\bibnamefont
  {Gutfleisch}}, \bibinfo {author} {\bibfnamefont {T.}~\bibnamefont
  {Gottschall}}, \bibinfo {author} {\bibfnamefont {M.}~\bibnamefont {Fries}},
  \bibinfo {author} {\bibfnamefont {D.}~\bibnamefont {Benke}}, \bibinfo
  {author} {\bibfnamefont {I.}~\bibnamefont {Radulov}}, \bibinfo {author}
  {\bibfnamefont {K.~P.}\ \bibnamefont {Skokov}}, \bibinfo {author}
  {\bibfnamefont {H.}~\bibnamefont {Wende}}, \bibinfo {author} {\bibfnamefont
  {M.}~\bibnamefont {Gruner}}, \bibinfo {author} {\bibfnamefont
  {M.}~\bibnamefont {Acet}}, \bibinfo {author} {\bibfnamefont {P.}~\bibnamefont
  {Entel}},\ and\ \bibinfo {author} {\bibfnamefont {M.}~\bibnamefont {Farle}},\
  }\href@noop {} {\bibfield  {journal} {\bibinfo  {journal} {Phil. Trans. R.
  Soc. A}\ }\textbf {\bibinfo {volume} {374}},\ \bibinfo {pages} {20150308}
  (\bibinfo {year} {2016})}\BibitemShut {NoStop}%
\bibitem [{\citenamefont {Nguyen}\ \emph {et~al.}(2011)\citenamefont {Nguyen},
  \citenamefont {Ou}, \citenamefont {Caron}, \citenamefont {Zhang},
  \citenamefont {Thanh}, \citenamefont {de~Wijs}, \citenamefont {de~Groot},
  \citenamefont {Buschow},\ and\ \citenamefont {Brueck}}]{NguyenAdvEn2011}%
  \BibitemOpen
  \bibfield  {author} {\bibinfo {author} {\bibfnamefont {H.~D.}\ \bibnamefont
  {Nguyen}}, \bibinfo {author} {\bibfnamefont {Z.~Q.}\ \bibnamefont {Ou}},
  \bibinfo {author} {\bibfnamefont {L.}~\bibnamefont {Caron}}, \bibinfo
  {author} {\bibfnamefont {L.}~\bibnamefont {Zhang}}, \bibinfo {author}
  {\bibfnamefont {D.~T.~C.}\ \bibnamefont {Thanh}}, \bibinfo {author}
  {\bibfnamefont {G.~A.}\ \bibnamefont {de~Wijs}}, \bibinfo {author}
  {\bibfnamefont {R.}~\bibnamefont {de~Groot}}, \bibinfo {author}
  {\bibfnamefont {K.~H.~J.}\ \bibnamefont {Buschow}},\ and\ \bibinfo {author}
  {\bibfnamefont {E.}~\bibnamefont {Brueck}},\ }\href@noop {} {\bibfield
  {journal} {\bibinfo  {journal} {Advanced Energy Materials}\ }\textbf
  {\bibinfo {volume} {1}},\ \bibinfo {pages} {1215} (\bibinfo {year}
  {2011})}\BibitemShut {NoStop}%
\bibitem [{\citenamefont {Waske}(2019)}]{WaskeNatEn2019}%
  \BibitemOpen
  \bibfield  {author} {\bibinfo {author} {\bibfnamefont {A.}~\bibnamefont
  {Waske}},\ }\href@noop {} {\bibfield  {journal} {\bibinfo  {journal} {Nature
  Energy}\ }\textbf {\bibinfo {volume} {4}},\ \bibinfo {pages} {68} (\bibinfo
  {year} {2019})}\BibitemShut {NoStop}%
\bibitem [{\citenamefont {Moore}\ \emph {et~al.}(2009)\citenamefont {Moore},
  \citenamefont {Morrison}, \citenamefont {Sandeman}, \citenamefont {Katter},\
  and\ \citenamefont {Cohen}}]{MooreExtrinsic}%
  \BibitemOpen
  \bibfield  {author} {\bibinfo {author} {\bibfnamefont {J.~D.}\ \bibnamefont
  {Moore}}, \bibinfo {author} {\bibfnamefont {K.}~\bibnamefont {Morrison}},
  \bibinfo {author} {\bibfnamefont {K.~G.}\ \bibnamefont {Sandeman}}, \bibinfo
  {author} {\bibfnamefont {M.}~\bibnamefont {Katter}},\ and\ \bibinfo {author}
  {\bibfnamefont {L.~F.}\ \bibnamefont {Cohen}},\ }\href@noop {} {\bibfield
  {journal} {\bibinfo  {journal} {Appl. Phys. Lett.}\ }\textbf {\bibinfo
  {volume} {95}},\ \bibinfo {pages} {252504} (\bibinfo {year}
  {2009})}\BibitemShut {NoStop}%
\bibitem [{\citenamefont {Morrison}\ and\ \citenamefont
  {Cohen}(2014)}]{MorrisonMetMat2014}%
  \BibitemOpen
  \bibfield  {author} {\bibinfo {author} {\bibfnamefont {K.}~\bibnamefont
  {Morrison}}\ and\ \bibinfo {author} {\bibfnamefont {L.~F.}\ \bibnamefont
  {Cohen}},\ }\href@noop {} {\bibfield  {journal} {\bibinfo  {journal} {Met.
  Mat. Trans. E}\ }\textbf {\bibinfo {volume} {1}},\ \bibinfo {pages} {153}
  (\bibinfo {year} {2014})}\BibitemShut {NoStop}%
\bibitem [{\citenamefont {Kuz'min}\ and\ \citenamefont
  {Richter}(2007)}]{KuzminPRB2007}%
  \BibitemOpen
  \bibfield  {author} {\bibinfo {author} {\bibfnamefont {M.~D.}\ \bibnamefont
  {Kuz'min}}\ and\ \bibinfo {author} {\bibfnamefont {M.}~\bibnamefont
  {Richter}},\ }\href@noop {} {\bibfield  {journal} {\bibinfo  {journal} {Phys.
  Rev. B}\ }\textbf {\bibinfo {volume} {76}},\ \bibinfo {pages} {092401}
  (\bibinfo {year} {2007})}\BibitemShut {NoStop}%
\bibitem [{\citenamefont {Fujita}\ \emph {et~al.}(2004)\citenamefont {Fujita},
  \citenamefont {Fujieda},\ and\ \citenamefont {Fukamichi}}]{FujitaJMMM2004}%
  \BibitemOpen
  \bibfield  {author} {\bibinfo {author} {\bibfnamefont {A.}~\bibnamefont
  {Fujita}}, \bibinfo {author} {\bibfnamefont {S.}~\bibnamefont {Fujieda}},\
  and\ \bibinfo {author} {\bibfnamefont {K.}~\bibnamefont {Fukamichi}},\
  }\href@noop {} {\bibfield  {journal} {\bibinfo  {journal} {J. Magn. Magn.
  Mater.}\ }\textbf {\bibinfo {volume} {272-276}},\ \bibinfo {pages} {e629}
  (\bibinfo {year} {2004})}\BibitemShut {NoStop}%
\bibitem [{\citenamefont {Morrison}\ \emph
  {et~al.}(2012{\natexlab{a}})\citenamefont {Morrison}, \citenamefont
  {Lyubina}, \citenamefont {Moore}, \citenamefont {Sandeman}, \citenamefont
  {Gutfleisch}, \citenamefont {Cohen},\ and\ \citenamefont
  {Caplin}}]{MorrisonPhilMag}%
  \BibitemOpen
  \bibfield  {author} {\bibinfo {author} {\bibfnamefont {K.}~\bibnamefont
  {Morrison}}, \bibinfo {author} {\bibfnamefont {J.}~\bibnamefont {Lyubina}},
  \bibinfo {author} {\bibfnamefont {J.~D.}\ \bibnamefont {Moore}}, \bibinfo
  {author} {\bibfnamefont {K.~G.}\ \bibnamefont {Sandeman}}, \bibinfo {author}
  {\bibfnamefont {O.}~\bibnamefont {Gutfleisch}}, \bibinfo {author}
  {\bibfnamefont {L.~F.}\ \bibnamefont {Cohen}},\ and\ \bibinfo {author}
  {\bibfnamefont {A.~D.}\ \bibnamefont {Caplin}},\ }\href@noop {} {\bibfield
  {journal} {\bibinfo  {journal} {Phil. Mag.}\ }\textbf {\bibinfo {volume}
  {92}},\ \bibinfo {pages} {292} (\bibinfo {year}
  {2012}{\natexlab{a}})}\BibitemShut {NoStop}%
\bibitem [{\citenamefont {Fujita}(2016)}]{FujitaAPL2016}%
  \BibitemOpen
  \bibfield  {author} {\bibinfo {author} {\bibfnamefont {A.}~\bibnamefont
  {Fujita}},\ }\href@noop {} {\bibfield  {journal} {\bibinfo  {journal} {APL
  Mater.}\ }\textbf {\bibinfo {volume} {4}},\ \bibinfo {pages} {064108}
  (\bibinfo {year} {2016})}\BibitemShut {NoStop}%
\bibitem [{\citenamefont {Faske}\ \emph {et~al.}(2020)\citenamefont {Faske},
  \citenamefont {Radulov}, \citenamefont {Hoelzel}, \citenamefont
  {Gutfleisch},\ and\ \citenamefont {Donner}}]{FaskeJPC2020}%
  \BibitemOpen
  \bibfield  {author} {\bibinfo {author} {\bibfnamefont {T.}~\bibnamefont
  {Faske}}, \bibinfo {author} {\bibfnamefont {I.~A.}\ \bibnamefont {Radulov}},
  \bibinfo {author} {\bibfnamefont {M.}~\bibnamefont {Hoelzel}}, \bibinfo
  {author} {\bibfnamefont {O.}~\bibnamefont {Gutfleisch}},\ and\ \bibinfo
  {author} {\bibfnamefont {W.}~\bibnamefont {Donner}},\ }\href@noop {}
  {\bibfield  {journal} {\bibinfo  {journal} {J Phys.:Condens. Matter}\
  }\textbf {\bibinfo {volume} {32}},\ \bibinfo {pages} {115802} (\bibinfo
  {year} {2020})}\BibitemShut {NoStop}%
\bibitem [{\citenamefont {Zhang}\ \emph {et~al.}(2021)\citenamefont {Zhang},
  \citenamefont {Zhou}, \citenamefont {Mole}, \citenamefont {Yu}, \citenamefont
  {Zhang}, \citenamefont {Zhao}, \citenamefont {Yu}, \citenamefont {Li},
  \citenamefont {Hu}, \citenamefont {Shen},\ and\ \citenamefont
  {Zhang}}]{Zhang0p5LFS}%
  \BibitemOpen
  \bibfield  {author} {\bibinfo {author} {\bibfnamefont {Z.}~\bibnamefont
  {Zhang}}, \bibinfo {author} {\bibfnamefont {H.}~\bibnamefont {Zhou}},
  \bibinfo {author} {\bibfnamefont {R.}~\bibnamefont {Mole}}, \bibinfo {author}
  {\bibfnamefont {C.}~\bibnamefont {Yu}}, \bibinfo {author} {\bibfnamefont
  {Z.}~\bibnamefont {Zhang}}, \bibinfo {author} {\bibfnamefont
  {W.}~\bibnamefont {Zhao}, \bibfnamefont {X.~Ren}}, \bibinfo {author}
  {\bibfnamefont {D.}~\bibnamefont {Yu}}, \bibinfo {author} {\bibfnamefont
  {B.}~\bibnamefont {Li}}, \bibinfo {author} {\bibfnamefont {F.}~\bibnamefont
  {Hu}}, \bibinfo {author} {\bibfnamefont {B.}~\bibnamefont {Shen}},\ and\
  \bibinfo {author} {\bibfnamefont {Z.}~\bibnamefont {Zhang}},\ }\href@noop {}
  {\bibfield  {journal} {\bibinfo  {journal} {Physical Review Materials}\
  }\textbf {\bibinfo {volume} {5}},\ \bibinfo {pages} {L071401} (\bibinfo
  {year} {2021})}\BibitemShut {NoStop}%
\bibitem [{INS()}]{INS}%
  \BibitemOpen
  \href@noop {} {\bibinfo  {journal} {INS measurements were performed on the
  MARI and LET time of flight direct geometry spectrometers at the ISIS Neutron
  and Muon Source, UK.}\ }\BibitemShut {NoStop}%
\bibitem [{\citenamefont {Bewley}\ \emph {et~al.}(2011)\citenamefont {Bewley},
  \citenamefont {Taylor},\ and\ \citenamefont {Bennington}}]{BewleyNucIns2011}%
  \BibitemOpen
\bibfield  {journal} {  }\bibfield  {author} {\bibinfo {author} {\bibfnamefont
  {R.~I.}\ \bibnamefont {Bewley}}, \bibinfo {author} {\bibfnamefont {J.~W.}\
  \bibnamefont {Taylor}},\ and\ \bibinfo {author} {\bibfnamefont {S.~M.}\
  \bibnamefont {Bennington}},\ }\href@noop {} {\bibfield  {journal} {\bibinfo
  {journal} {Nuclear Instruments and Methods in Physics}\ }\textbf {\bibinfo
  {volume} {637}},\ \bibinfo {pages} {128} (\bibinfo {year}
  {2011})}\BibitemShut {NoStop}%
\bibitem [{\citenamefont {Rodriguez-Carvajal}(1993)}]{Fullprof}%
  \BibitemOpen
  \bibfield  {author} {\bibinfo {author} {\bibfnamefont {J.}~\bibnamefont
  {Rodriguez-Carvajal}},\ }\href@noop {} {\bibfield  {journal} {\bibinfo
  {journal} {Physica B}\ }\textbf {\bibinfo {volume} {192}},\ \bibinfo {pages}
  {55} (\bibinfo {year} {1993})}\BibitemShut {NoStop}%
\bibitem [{\citenamefont {Minakov}\ \emph {et~al.}(2005)\citenamefont
  {Minakov}, \citenamefont {Roy}, \citenamefont {Bugoslavsky},\ and\
  \citenamefont {Cohen}}]{Minakov}%
  \BibitemOpen
  \bibfield  {author} {\bibinfo {author} {\bibfnamefont {A.~A.}\ \bibnamefont
  {Minakov}}, \bibinfo {author} {\bibfnamefont {S.~B.}\ \bibnamefont {Roy}},
  \bibinfo {author} {\bibfnamefont {Y.~V.}\ \bibnamefont {Bugoslavsky}},\ and\
  \bibinfo {author} {\bibfnamefont {L.~F.}\ \bibnamefont {Cohen}},\ }\href@noop
  {} {\bibfield  {journal} {\bibinfo  {journal} {Rev. Sci. Instrum.}\ }\textbf
  {\bibinfo {volume} {76}},\ \bibinfo {pages} {043906} (\bibinfo {year}
  {2005})}\BibitemShut {NoStop}%
\bibitem [{\citenamefont {Miyoshi}\ \emph {et~al.}(2008)\citenamefont
  {Miyoshi}, \citenamefont {Morrison}, \citenamefont {Moore}, \citenamefont
  {Caplin},\ and\ \citenamefont {Cohen}}]{MiyoshiLH}%
  \BibitemOpen
  \bibfield  {author} {\bibinfo {author} {\bibfnamefont {Y.}~\bibnamefont
  {Miyoshi}}, \bibinfo {author} {\bibfnamefont {K.}~\bibnamefont {Morrison}},
  \bibinfo {author} {\bibfnamefont {J.~D.}\ \bibnamefont {Moore}}, \bibinfo
  {author} {\bibfnamefont {A.~D.}\ \bibnamefont {Caplin}},\ and\ \bibinfo
  {author} {\bibfnamefont {L.~F.}\ \bibnamefont {Cohen}},\ }\href@noop {}
  {\bibfield  {journal} {\bibinfo  {journal} {Rev. Sci. Instrum.}\ }\textbf
  {\bibinfo {volume} {79}},\ \bibinfo {pages} {074901} (\bibinfo {year}
  {2008})}\BibitemShut {NoStop}%
\bibitem [{\citenamefont {Morrison}\ \emph {et~al.}(2010)\citenamefont
  {Morrison}, \citenamefont {Lyubina}, \citenamefont {Moore}, \citenamefont
  {Caplin}, \citenamefont {Sandeman}, \citenamefont {Gutfleisch},\ and\
  \citenamefont {Cohen}}]{MorrisonJPD2010}%
  \BibitemOpen
  \bibfield  {author} {\bibinfo {author} {\bibfnamefont {K.}~\bibnamefont
  {Morrison}}, \bibinfo {author} {\bibfnamefont {J.}~\bibnamefont {Lyubina}},
  \bibinfo {author} {\bibfnamefont {J.~D.}\ \bibnamefont {Moore}}, \bibinfo
  {author} {\bibfnamefont {A.~D.}\ \bibnamefont {Caplin}}, \bibinfo {author}
  {\bibfnamefont {K.~G.}\ \bibnamefont {Sandeman}}, \bibinfo {author}
  {\bibfnamefont {O.}~\bibnamefont {Gutfleisch}},\ and\ \bibinfo {author}
  {\bibfnamefont {L.~F.}\ \bibnamefont {Cohen}},\ }\href@noop {} {\bibfield
  {journal} {\bibinfo  {journal} {J. Phys. D: Appl. Phys.}\ }\textbf {\bibinfo
  {volume} {43}},\ \bibinfo {pages} {132001} (\bibinfo {year}
  {2010})}\BibitemShut {NoStop}%
\bibitem [{\citenamefont {Morrison}\ \emph
  {et~al.}(2012{\natexlab{b}})\citenamefont {Morrison}, \citenamefont {Bratko},
  \citenamefont {Turcaud}, \citenamefont {Berenov}, \citenamefont {Caplin},\
  and\ \citenamefont {Cohen}}]{MorrisonLHSO}%
  \BibitemOpen
  \bibfield  {author} {\bibinfo {author} {\bibfnamefont {K.}~\bibnamefont
  {Morrison}}, \bibinfo {author} {\bibfnamefont {M.}~\bibnamefont {Bratko}},
  \bibinfo {author} {\bibfnamefont {J.}~\bibnamefont {Turcaud}}, \bibinfo
  {author} {\bibfnamefont {A.}~\bibnamefont {Berenov}}, \bibinfo {author}
  {\bibfnamefont {A.~D.}\ \bibnamefont {Caplin}},\ and\ \bibinfo {author}
  {\bibfnamefont {L.~F.}\ \bibnamefont {Cohen}},\ }\href@noop {} {\bibfield
  {journal} {\bibinfo  {journal} {Rev. Sci. Instrum.}\ }\textbf {\bibinfo
  {volume} {83}},\ \bibinfo {pages} {033901} (\bibinfo {year}
  {2012}{\natexlab{b}})}\BibitemShut {NoStop}%
\bibitem [{\citenamefont {Morrison}\ \emph {et~al.}(2009)\citenamefont
  {Morrison}, \citenamefont {Moore}, \citenamefont {Sandeman}, \citenamefont
  {Caplin},\ and\ \citenamefont {Cohen}}]{MorrisonPRB2009}%
  \BibitemOpen
  \bibfield  {author} {\bibinfo {author} {\bibfnamefont {K.}~\bibnamefont
  {Morrison}}, \bibinfo {author} {\bibfnamefont {J.~D.}\ \bibnamefont {Moore}},
  \bibinfo {author} {\bibfnamefont {K.~G.}\ \bibnamefont {Sandeman}}, \bibinfo
  {author} {\bibfnamefont {A.~D.}\ \bibnamefont {Caplin}},\ and\ \bibinfo
  {author} {\bibfnamefont {L.~F.}\ \bibnamefont {Cohen}},\ }\href@noop {}
  {\bibfield  {journal} {\bibinfo  {journal} {Phys. Rev. B}\ }\textbf {\bibinfo
  {volume} {79}},\ \bibinfo {pages} {134408} (\bibinfo {year}
  {2009})}\BibitemShut {NoStop}%
\bibitem [{\citenamefont {Morrison}\ \emph {et~al.}(2013)\citenamefont
  {Morrison}, \citenamefont {Dupas}, \citenamefont {Mudryk}, \citenamefont
  {Pecharsky}, \citenamefont {Gschneidner}, \citenamefont {Caplin},\ and\
  \citenamefont {Cohen}}]{MorrisonPRB2013}%
  \BibitemOpen
  \bibfield  {author} {\bibinfo {author} {\bibfnamefont {K.}~\bibnamefont
  {Morrison}}, \bibinfo {author} {\bibfnamefont {A.}~\bibnamefont {Dupas}},
  \bibinfo {author} {\bibfnamefont {Y.}~\bibnamefont {Mudryk}}, \bibinfo
  {author} {\bibfnamefont {V.~K.}\ \bibnamefont {Pecharsky}}, \bibinfo {author}
  {\bibfnamefont {K.~A.}\ \bibnamefont {Gschneidner}}, \bibinfo {author}
  {\bibfnamefont {A.~D.}\ \bibnamefont {Caplin}},\ and\ \bibinfo {author}
  {\bibfnamefont {L.~F.}\ \bibnamefont {Cohen}},\ }\href@noop {} {\bibfield
  {journal} {\bibinfo  {journal} {Phys. Rev. B}\ }\textbf {\bibinfo {volume}
  {87}},\ \bibinfo {pages} {134421} (\bibinfo {year} {2013})}\BibitemShut
  {NoStop}%
\bibitem [{\citenamefont {Morrison}\ \emph {et~al.}(2011)\citenamefont
  {Morrison}, \citenamefont {Podgornykn}, \citenamefont {Shcherbakova},
  \citenamefont {Caplin},\ and\ \citenamefont {Cohen}}]{MorrisonPRB2011}%
  \BibitemOpen
  \bibfield  {author} {\bibinfo {author} {\bibfnamefont {K.}~\bibnamefont
  {Morrison}}, \bibinfo {author} {\bibfnamefont {S.~M.}\ \bibnamefont
  {Podgornykn}}, \bibinfo {author} {\bibfnamefont {Y.~V.}\ \bibnamefont
  {Shcherbakova}}, \bibinfo {author} {\bibfnamefont {A.~D.}\ \bibnamefont
  {Caplin}},\ and\ \bibinfo {author} {\bibfnamefont {L.~F.}\ \bibnamefont
  {Cohen}},\ }\href@noop {} {\bibfield  {journal} {\bibinfo  {journal} {Phys.
  Rev. B}\ }\textbf {\bibinfo {volume} {83}},\ \bibinfo {pages} {144415}
  (\bibinfo {year} {2011})}\BibitemShut {NoStop}%
\bibitem [{\citenamefont {Morrison}()}]{MorrisonPRB2021}%
  \BibitemOpen
  \bibfield  {author} {\bibinfo {author} {\bibfnamefont {K.}~\bibnamefont
  {Morrison}},\ }\href@noop {} {\bibinfo  {journal} {to be submitted to
  Physical Review B}\ }\BibitemShut {NoStop}%
\bibitem [{\citenamefont {Gruner}\ \emph {et~al.}(2018)\citenamefont {Gruner},
  \citenamefont {Keune}, \citenamefont {Landers}, \citenamefont {Salamon},
  \citenamefont {Krautz}, \citenamefont {Zhao}, \citenamefont {Hu},
  \citenamefont {Toellner}, \citenamefont {Alp}, \citenamefont {Gutfleisch},\
  and\ \citenamefont {Wende}}]{GrunerPSS}%
  \BibitemOpen
\bibfield  {journal} {  }\bibfield  {author} {\bibinfo {author} {\bibfnamefont
  {M.~E.}\ \bibnamefont {Gruner}}, \bibinfo {author} {\bibfnamefont
  {W.}~\bibnamefont {Keune}}, \bibinfo {author} {\bibfnamefont
  {J.}~\bibnamefont {Landers}}, \bibinfo {author} {\bibfnamefont
  {S.}~\bibnamefont {Salamon}}, \bibinfo {author} {\bibfnamefont
  {M.}~\bibnamefont {Krautz}}, \bibinfo {author} {\bibfnamefont
  {J.}~\bibnamefont {Zhao}}, \bibinfo {author} {\bibfnamefont {M.~Y.}\
  \bibnamefont {Hu}}, \bibinfo {author} {\bibfnamefont {T.}~\bibnamefont
  {Toellner}}, \bibinfo {author} {\bibfnamefont {E.~E.}\ \bibnamefont {Alp}},
  \bibinfo {author} {\bibfnamefont {O.}~\bibnamefont {Gutfleisch}},\ and\
  \bibinfo {author} {\bibfnamefont {H.}~\bibnamefont {Wende}},\ }\href@noop {}
  {\bibfield  {journal} {\bibinfo  {journal} {Phys. Status Solidi B}\ }\textbf
  {\bibinfo {volume} {255}},\ \bibinfo {pages} {1700465} (\bibinfo {year}
  {2018})}\BibitemShut {NoStop}%
\bibitem [{\citenamefont {Landers}\ \emph {et~al.}(2018)\citenamefont
  {Landers}, \citenamefont {Salamon}, \citenamefont {Keune}, \citenamefont
  {Gruner}, \citenamefont {Krautz}, \citenamefont {Zhao}, \citenamefont {Hu},
  \citenamefont {Toellner}, \citenamefont {Alp}, \citenamefont {Gutfleisch},\
  and\ \citenamefont {Wende}}]{LanderPRB2018}%
  \BibitemOpen
  \bibfield  {author} {\bibinfo {author} {\bibfnamefont {J.}~\bibnamefont
  {Landers}}, \bibinfo {author} {\bibfnamefont {S.}~\bibnamefont {Salamon}},
  \bibinfo {author} {\bibfnamefont {W.}~\bibnamefont {Keune}}, \bibinfo
  {author} {\bibfnamefont {M.~E.}\ \bibnamefont {Gruner}}, \bibinfo {author}
  {\bibfnamefont {M.}~\bibnamefont {Krautz}}, \bibinfo {author} {\bibfnamefont
  {J.}~\bibnamefont {Zhao}}, \bibinfo {author} {\bibfnamefont {M.~Y.}\
  \bibnamefont {Hu}}, \bibinfo {author} {\bibfnamefont {T.}~\bibnamefont
  {Toellner}}, \bibinfo {author} {\bibfnamefont {E.~E.}\ \bibnamefont {Alp}},
  \bibinfo {author} {\bibfnamefont {O.}~\bibnamefont {Gutfleisch}},\ and\
  \bibinfo {author} {\bibfnamefont {H.}~\bibnamefont {Wende}},\ }\href@noop {}
  {\bibfield  {journal} {\bibinfo  {journal} {Physical Review B}\ }\textbf
  {\bibinfo {volume} {98}},\ \bibinfo {pages} {024417} (\bibinfo {year}
  {2018})}\BibitemShut {NoStop}%
\bibitem [{\citenamefont {Berridge}\ \emph {et~al.}(2010)\citenamefont
  {Berridge}, \citenamefont {Grigera}, \citenamefont {Simons},\ and\
  \citenamefont {Green}}]{Berridge-PRB-2010}%
  \BibitemOpen
  \bibfield  {author} {\bibinfo {author} {\bibfnamefont {A.~M.}\ \bibnamefont
  {Berridge}}, \bibinfo {author} {\bibfnamefont {S.~A.}\ \bibnamefont
  {Grigera}}, \bibinfo {author} {\bibfnamefont {B.~D.}\ \bibnamefont
  {Simons}},\ and\ \bibinfo {author} {\bibfnamefont {A.~G.}\ \bibnamefont
  {Green}},\ }\href@noop {} {\bibfield  {journal} {\bibinfo  {journal} {Phys.
  Rev. B}\ }\textbf {\bibinfo {volume} {81}},\ \bibinfo {pages} {054429}
  (\bibinfo {year} {2010})}\BibitemShut {NoStop}%
\bibitem [{\citenamefont {Efremov}\ \emph {et~al.}(2019)\citenamefont
  {Efremov}, \citenamefont {Shtyk}, \citenamefont {Rost}, \citenamefont
  {Chamon}, \citenamefont {Mackenzie},\ and\ \citenamefont
  {Betouras}}]{Betouras-PRL-2019}%
  \BibitemOpen
  \bibfield  {author} {\bibinfo {author} {\bibfnamefont {D.~V.}\ \bibnamefont
  {Efremov}}, \bibinfo {author} {\bibfnamefont {A.}~\bibnamefont {Shtyk}},
  \bibinfo {author} {\bibfnamefont {A.~W.}\ \bibnamefont {Rost}}, \bibinfo
  {author} {\bibfnamefont {C.}~\bibnamefont {Chamon}}, \bibinfo {author}
  {\bibfnamefont {A.~P.}\ \bibnamefont {Mackenzie}},\ and\ \bibinfo {author}
  {\bibfnamefont {J.~J.}\ \bibnamefont {Betouras}},\ }\href@noop {} {\bibfield
  {journal} {\bibinfo  {journal} {Phys. Rev. Lett.}\ }\textbf {\bibinfo
  {volume} {113}},\ \bibinfo {pages} {207202} (\bibinfo {year}
  {2019})}\BibitemShut {NoStop}%
\bibitem [{\citenamefont {Belitz}\ \emph {et~al.}(1999)\citenamefont {Belitz},
  \citenamefont {Kirkpatrick},\ and\ \citenamefont {Vojta}}]{Belitz-PRL-1999}%
  \BibitemOpen
  \bibfield  {author} {\bibinfo {author} {\bibfnamefont {D.}~\bibnamefont
  {Belitz}}, \bibinfo {author} {\bibfnamefont {T.~R.}\ \bibnamefont
  {Kirkpatrick}},\ and\ \bibinfo {author} {\bibfnamefont {T.}~\bibnamefont
  {Vojta}},\ }\href@noop {} {\bibfield  {journal} {\bibinfo  {journal} {Phys.
  Rev. B}\ }\textbf {\bibinfo {volume} {82}},\ \bibinfo {pages} {4707}
  (\bibinfo {year} {1999})}\BibitemShut {NoStop}%
\bibitem [{\citenamefont {Chubukov}\ \emph {et~al.}(2004)\citenamefont
  {Chubukov}, \citenamefont {Pepin},\ and\ \citenamefont
  {Rech}}]{Chubukov-PRL-2004}%
  \BibitemOpen
  \bibfield  {author} {\bibinfo {author} {\bibfnamefont {A.~V.}\ \bibnamefont
  {Chubukov}}, \bibinfo {author} {\bibfnamefont {C.}~\bibnamefont {Pepin}},\
  and\ \bibinfo {author} {\bibfnamefont {J.}~\bibnamefont {Rech}},\ }\href@noop
  {} {\bibfield  {journal} {\bibinfo  {journal} {Phys. Rev. Lett.}\ }\textbf
  {\bibinfo {volume} {92}},\ \bibinfo {pages} {147003} (\bibinfo {year}
  {2004})}\BibitemShut {NoStop}%
\bibitem [{\citenamefont {Efremov}\ \emph {et~al.}(2008)\citenamefont
  {Efremov}, \citenamefont {Betouras},\ and\ \citenamefont
  {Chubukov}}]{Betouras-PRB-2008}%
  \BibitemOpen
  \bibfield  {author} {\bibinfo {author} {\bibfnamefont {D.~V.}\ \bibnamefont
  {Efremov}}, \bibinfo {author} {\bibfnamefont {J.~J.}\ \bibnamefont
  {Betouras}},\ and\ \bibinfo {author} {\bibfnamefont {A.~V.}\ \bibnamefont
  {Chubukov}},\ }\href@noop {} {\bibfield  {journal} {\bibinfo  {journal}
  {Phys. Rev. B}\ }\textbf {\bibinfo {volume} {77}},\ \bibinfo {pages}
  {220401(R)} (\bibinfo {year} {2008})}\BibitemShut {NoStop}%
\bibitem [{\citenamefont {Karahasanovic}\ \emph {et~al.}(2012)\citenamefont
  {Karahasanovic}, \citenamefont {Krüger},\ and\ \citenamefont
  {Green}}]{Karahasanovic-PRB-2012}%
  \BibitemOpen
  \bibfield  {author} {\bibinfo {author} {\bibfnamefont {U.}~\bibnamefont
  {Karahasanovic}}, \bibinfo {author} {\bibfnamefont {F.}~\bibnamefont
  {Krüger}},\ and\ \bibinfo {author} {\bibfnamefont {A.~G.}\ \bibnamefont
  {Green}},\ }\href@noop {} {\bibfield  {journal} {\bibinfo  {journal} {Phys.
  Rev. B}\ }\textbf {\bibinfo {volume} {85}},\ \bibinfo {pages} {165111}
  (\bibinfo {year} {2012})}\BibitemShut {NoStop}%
\bibitem [{\citenamefont {Brando}\ \emph {et~al.}(2008)\citenamefont {Brando},
  \citenamefont {Belitz}, \citenamefont {Grosche},\ and\ \citenamefont
  {Kirkpatrick}}]{Brando-RMP}%
  \BibitemOpen
  \bibfield  {author} {\bibinfo {author} {\bibfnamefont {M.}~\bibnamefont
  {Brando}}, \bibinfo {author} {\bibfnamefont {D.}~\bibnamefont {Belitz}},
  \bibinfo {author} {\bibfnamefont {F.~M.}\ \bibnamefont {Grosche}},\ and\
  \bibinfo {author} {\bibfnamefont {T.~R.}\ \bibnamefont {Kirkpatrick}},\
  }\href@noop {} {\bibfield  {journal} {\bibinfo  {journal} {Rev. Mod. Phys.}\
  }\textbf {\bibinfo {volume} {77}},\ \bibinfo {pages} {220401(R)} (\bibinfo
  {year} {2008})}\BibitemShut {NoStop}%
\end{thebibliography}%
